\documentclass[12pt, letter]{article}

\usepackage{xr-hyper}       

\usepackage{graphicx}
\usepackage{enumerate}
\usepackage{natbib}

\usepackage{amsmath, amsfonts, amssymb, amsthm}
\usepackage{booktabs}       
\usepackage{multirow}
\usepackage{sectsty}
\usepackage{tabularx}
\usepackage{graphicx}
\usepackage{bm}             
\usepackage{xcolor}
\usepackage{microtype} 
\usepackage{import}

\usepackage{url}

\usepackage{breakurl}
\usepackage[breaklinks]{hyperref}

\usepackage{soul, color}
\usepackage{titling}

\graphicspath{{./figures/}}

\newcommand{\dblink}{\texttt{\upshape \lowercase{d-blink}}} 
\newcommand{\blink}{\texttt{\upshape \lowercase{blink}}} 

\newcommand{\eat}[1]{}

\definecolor{mblue}{rgb}{0,0.4470,0.7410}
\hypersetup{pdfborder={0 0 0},colorlinks=true,linkcolor=mblue,urlcolor=mblue,citecolor=mblue}
\usepackage[utf8]{inputenc}
\usepackage{amsmath,amsfonts,amssymb,amsthm}
\usepackage{authblk}
\usepackage{comment}
\usepackage{soul}

\newcommand{\ob}[1]{\textcolor{blue}{#1}}

\newcommand{\blind}{1}

\usepackage[margin=1in]{geometry}

\begin{document}

\def\spacingset#1{\renewcommand{\baselinestretch}%
{#1}\small\normalsize} \spacingset{1}


\if1\blind
{
  \title{\bf (Almost) All of Entity Resolution}
  \author{Olivier Binette and Rebecca C. Steorts\footnote{Olivier Binette is a PhD Student in the Department of Statistical Science at Duke University, Durham NC 27708 (e-mail: \url{olivier.binette@duke.edu}). Rebecca C. Steorts is Assistant Professor of Statistical Science and affiliated faculty in Computer Science, Biostatistics and Bioinformatics, the information initiative at Duke (iiD) and the Social Science Research Institute (SSRI), Duke University, Durham NC 27708, and Principal Mathematical Statistician at the U.S. Census Bureau (e-mail: \url{beka@stat.duke.edu}). The authors gratefully acknowledge the support of the Natural Sciences and Engineering Research Council of Canada, of the Fonds de Recherche du Québec --- Nature et Technologies, of NSF Career Award 1652431 and of the Alfred P. Sloan Foundation. The authors thank Ted Enamorado and MJM for providing comments that greatly improved this work.}\\\hspace{.2cm}
    Duke University\\ Durham, NC 27708}
  \maketitle
} \fi

\if0\blind
{
  \bigskip
  \bigskip
  \bigskip
  \begin{center}
    {\LARGE\bf (Almost) All of Entity Resolution}
\end{center} 
  \medskip
} \fi

\bigskip

\begin{abstract}
    Whether the goal is to estimate the number of people that live in a congressional district, to estimate the number of individuals that have died in an armed conflict, or to disambiguate individual authors using bibliographic data, all these applications have a common theme --- integrating information from multiple sources.  Before such questions can be answered, databases must be cleaned and integrated in a systematic and accurate way, commonly known as record linkage, de-duplication, or entity resolution. In this article, we review motivational applications and seminal papers that have led to the growth of this area. Specifically, we review the foundational work that began in the 1940's and 50's that have led to modern probabilistic record linkage. We review clustering approaches to entity resolution, semi- and fully supervised methods, and canonicalization, which are being used throughout industry and academia in applications such as human rights, official statistics, medicine, citation networks, among others. Finally, we discuss current research topics of practical importance.
\end{abstract}

\noindent%
{\it Keywords:} entity resolution, record linkage, de-duplication, clustering, human rights, official statistics, privacy

\spacingset{1.2} 

\section{Introduction}
\label{sec:introduction}

Information about individuals is often scattered across multiple databases. Combining such information can result in enormous benefits for analysis, resulting in richer and more reliable conclusions. In many applications, however, analysts cannot merge databases based on unique identifiers, often due to privacy reasons or due to the fact that they do not exist. Thus, analysts turn to methods from statistics, computer science, and machine learning known as {entity resolution}, record linkage, or deduplication.

Most entity resolution methods are motivated by applications that require the integration of databases before further analyses can occur. Such applications include the United States (U.S.) decennial census, casualty estimation in armed conflicts, voter registration data, and the analysis of co-authorship networks.
For example, in terms of the U.S. Census, it is important that there is a correct and accurate enumeration not only for apportioning the representation of legislators, but also for allocating resources for housing, highways, schools, assistance programs, and other projects that are vital to the prosperity, welfare, and economic growth in the U.S. Record linkage and de-duplication are used to ensure that everyone is counted once and only once. In the context of armed conflicts, enumerating the number of documented identifiable deaths in a conflict is relevant to the prosecution of war crimes and for justice around the world. This requires integrating and de-duplicating information about victims collected by multiple organizations. In terms of voter registration, it is critical that databases have duplicates removed from them such that voters have one unique voter registration record, and thus, vote only once. In addition, it is critical to know how many (unique) individuals are voting in elections and how many individuals are being left out, especially for minority groups. In order to study science and innovation at the individual level, authors and inventors in bibliographic databases must be disambiguated in the absence of unique identifiers.

As noted, the applications are widespread across many disciplines including statistics, computer science, economics, medicine, and human rights, among others. {Furthermore,} entity resolution is not only a crucial task for social science and industrial applications, but is also a challenging statistical and computational problem itself. This was recognized as early as in 1959 \citep{newcombe_automatic_1959} and the subject has since received growing interest from many communities with the rise of ``big data'' and modern computing.

This review article provides an introduction to entity resolution and its associated challenges. It overviews modern developments motivated by {the} social science applications {discussed} in Section~\ref{sec:motivational}. 
To our knowledge, other reviews of record linkage and entity resolution \citep{Gu2003, Brizan2006, elmagarmid_duplicate_2007, Herzog_2007, christen_data_2012, Winkler2014, dong_big_2015, Sayers2016, Christophides2019, Jurek2019} do not cover recent developments in Bayesian Fellegi-Sunter methodology (Section~\ref{sec:BayesianFS}), model-based clustering approaches (Section~\ref{sec:clust_model_based}), microclustering (Sections~\ref{sec:microclustering} and \ref{sec:feasibility}), and joint downstream tasks (Section~\ref{subsec:joint}). These principled statistical approaches to entity resolution provide uncertainty quantification and are particularly relevant to many social science applications. Our goal is to cover ``(almost) all of entity resolution" in this article that will reach a breadth of audiences and communities.

\subsection{Motivational Applications}\label{sec:motivational}

This section reviews social science applications that have motivated major developments in entity resolution. First, we introduce the U.S. decennial census which led to the proposed work of \cite{Jaro1989, winkler1991application, larsen_2001, winkler_overview_2006,  hogan2013quality, marchant_distributed_2019}. Second, we introduce an application to document identifiable deaths in El Salvador, which has led to many improvements in Bayesian ER methodology \citep{hoover2011repertoires, green2019civilian, sadinle_detecting_2014, sadinle_bayesian_2017, sadinle2018bayesian}. Third, we provide an application regarding voter registration in North Carolina, which has also led to developments in entity resolution as well as open-source software and data sets \citep{christen_preparation_2014, kaplan}. Finally, we discuss duplicated data that comes from publications or patent databases \citep{west2008pseudocode, treeratpituk2009disambiguating, ventura_2012, ventura2015seeing}.

\paragraph{Decennial Census.}
One critical problem in the U.S. occurs every ten years, when the U.S. Census Bureau attempts to enumerate the population as mandated under the U.S. Constitution, Article I, Section 2. An accurate enumeration is essential as this 
is used not only to apportion the representation of legislators, but also to allocate resources for housing, highways, schools, assistance programs, and other projects that are vital to the prosperity, welfare, and economic growth of the country.  As the U.S. grows and becomes more diverse, enumerating its population becomes more challenging. For example, many individuals elect not to fill out census forms due to privacy reasons, which results in these individuals not being represented in the enumeration. 
Other individuals are duplicated, e.g., students attending universities or private schools (living in group quarters) are often counted twice as they are legally required to be counted by their university/school, while also being counted by their parents/guardians as part of a household \citep{hogan2013quality}.

\paragraph{Documented Identifiable Deaths in El Salvador.}
Turning to the context of an armed conflict, creating  models enumerating identifiable deaths is challenging as grassroots movements, families, and friends collect multiple reports on the same victims. Between 1980 and 1991, the Republic of El Salvador witnessed a civil war between the central government, the left-wing guerrilla Farabundo Marti National Liberation Front (FMLN), and right-wing paramilitary death squads. After the peace agreement in 1992, the United Nations created a Commission on the Truth (UNTC) for El Salvador, which invited members of Salvadoran society to report war-related human rights violations, which mainly focused on killings and disappearances. In order to collect such information the UNTC invited individuals through newspapers, radio, and television advertisements to come forward and testify. The UNTC opened offices through El Salvador where witnesses could provide their testimonials, and this resulted in a list of potential victims with names, date of death, and reported location. 
Due to the fact that testimonials were provided to the UNTC many years after the civil war, it is expected that witnesses could not recall some of the details of the killings. In addition, some details regarding testimonials of the same individual may contain conflicting or differing information. This is a natural characteristic of this data and leads to more noise, distortions, and missingness. Furthermore, a victim can be reported multiple times, which leads to an issue with duplication in the data. Finally, 
there are no unique identifiers available for the majority of this data set \citep{hoover2011repertoires, green2019civilian, sadinle_detecting_2014}.

\paragraph{Estimation of Voters in North Carolina.}
While working with personally identifiable information is useful for understanding entity resolution models, such data are not typically made public due to privacy concerns and are instead stored in government databases where they cannot be published. One exception is that of voter registration databases, which are publicly available (and often online) in the United States. North Carolina makes its voter registration database freely and publicly available through the North Carolina State Board of Elections (NCSBE, \url{http://www.ncsbe.gov/}), which provides regular updates. This data set contains rich information, such as first and last name, year of birth, phone number, and address.  However, the voter registration number is often duplicated due to people moving, getting married, and various other reasons. See Table~\ref{tab:ex-records} for examples of public records from this data set.

The process through which voter registration records are matched with other official records can have a profound influence on one's ability to vote. Georgia's controversial ``exact match'' law \citep{ReutersGeorgia}, which was slightly changed in 2019, required an exact match between voter registration records and records from the Department of Driver Services or the Social Security Administration {in order to validate voter registrations}. For instance, typographical errors, different spellings of the same name, or outdated records could place a voter registration on hold. In 2017, about 670,000 registrations were canceled as a result \citep{APgeorgia}. \cite{TedGeorgia} showed how this law could predominantly affect non-white voters (see also \cite{PeoplevKemp}). 

\begin{table*}

\caption{\label{tab:ex-records} Example of public NCSBE records retrieved from \url{http://www.ncsbe.gov} for the county of Durham in North Carolina and corresponding to unique voter registration numbers. Some fields have been omitted for brevity, including ZIP code, phone number and voter registration number. Street addresses have been permuted with other individuals as to preserve some anonymity.}
\centering

\begin{tabular}[t]{ccccccc}
\toprule
Name  & Street Address  & Age & Sex & Race & Birth & Party \\
\midrule
Domineck Q. AAshad Jr & 914 Monmouth Ave \#3  & 26 & M & B & -- & LIB\\
Domineck Q. AAshad Sr & 1408 Auburndale Dr  & 55 & M & B & NY & DEM\\
Xiomara A. Martinez & 1715 Cole Mill Rd  & 31 & F & O & HL & REP\\
Xiomara A. Martinez & 2923 Forrestal Dr & 31 & F & O & HL & --\\
Virginia, L. Mullinix & 749 Ninth St \#480 & 101 & F & W & PA & REP\\
Jacqueline D. Fuller & 141 Bagby LN & 54 &  -- & -- & -- & DEM\\
Jacqueline Fuller & 905 Cook Rd & 56 &  F & B & NC & DEM\\

\bottomrule
\end{tabular}
\end{table*}

\paragraph{Inventor and Author Disambiguation.}
Entity resolution is used not just to remove duplication but also to enrich existing data. For instance, the United States Patent and Trademark Office (USPTO) maintains a patents database but does not provide unique identifiers for inventors. 
An inventor's affiliation commonly changes over time, and an inventor's name may contain distortions or may be duplicated. These two inconsistencies make linkage even more challenging. {For instance,} in the MEDLINE\footnote{MEDLINE is the U.S. National Library of Medicine bibliographic database that contains more than 26 million references to journal articles in life sciences with a concentration on biomedicine bibliographic databases.} database, \cite{torvik_2009} identified 15,980 publications {among 15.3 million} co-authored by ``J. Lee'' at the time. Furthermore, over two thirds of all authors were found to have a name which is shared by at least one other author. Resolving the individual inventors who have authored multiple patents or articles is necessary to build co-authorship networks, to track mobility across organizations, and to support research into the drivers of innovation \citep{Li2014}.

\subsection{Terminology}
The statistical and computational tools used to identify related records, remove duplicated entries, and aggregate information, have been referred to as entity resolution, record linkage, data matching, instance matching, data linkage, data cleaning, data fusion and merging \citep{christen_data_2012, dong_big_2015, Christophides2019}. Here, we introduce terminology which we use throughout the paper and which highlights different aspects of these processes.

First, \textit{databases} or \textit{files} are collections of \textit{records} which refer to \textit{entities} (such as a person, an object or an event). Each record contains information listed under a set of \textit{attributes}, \textit{fields} or \textit{features}, such as the person's first name, last name, gender, date of birth, etc. An attribute is a \textit{unique identifier} if two records refer to the same entity exactly when the unique identifiers are the same (for example, a birth number).

\textit{Deduplication} (or duplicate detection) refers to removing duplicate records within one database. This applies to census data, for example, where individuals 
may have been recorded multiple times (at different addresses).
\textit{Record linkage} is broadly used in the literature and refers to merging together multiple databases and removing duplicate records across the databases. 
{More precisely, \textit{bipartite record linkage} or \textit{maximum one-to-one} linkage refers to merging multiple databases and removing duplicate records across the databases, but not within the databases.} For example, we discuss in Section \ref{sec:probabilistic} the problem of linking birth records to marriage records which appeared in \cite{newcombe_automatic_1959}. Birth records in a first database were related to at most one marriage records in a second database, and duplicate birth or marriage records were not expected.
\textit{Entity resolution} refers to simultaneously merging together multiple databases and removing duplicate records across and within databases. 
This is relevant in the context of the documented deaths in El Salvador.
 Information about deaths are recorded by multiple organizations, which may result in a death being documented more than once through multiple organizations.
 

Given multiple records referring to the same entity, \textit{canonicalization} (data merging and data fusion) is the task of constructing one representative record after entity resolution to resolve any potential conflicting information. This is discussed in Section~\ref{sec:canonicalization}.

\subsection{Challenges of Entity Resolution}

Entity resolution is difficult because of the need to balance between: (1) efficient methods which scale to large databases, (2) accurate, robust and generalizable methods which make maximal use of all available information, and (3) appropriately quantifying and propagating uncertainty coming from all stages of the entity resolution pipeline. 

In the context of $k$ databases each with $N$ records, evaluating all $k$-tuples for possible links scales as $N^k$ in the number of records. Considering all record pairs, or clustering records referring to the same entities, typically scales as $N^2$. In practice, we seek methods which are subquadratic in the number of records. Computational speedups for this are known as \textit{blocking}. This is briefly discussed in Section \ref{sec:pipeline}, but it is not a focus of this review article. We refer the reader to \cite{Christophides2019, OHare2019, steorts_comparison_2014, murray2016probabilistic} for a more detailed review.

In addition, one seeks
methods which adapt to the data at hand, and which do not necessarily require training data as it is usually not available. These methods should generalize to a wide variety of applications. In addition, they should be robust to noise and distortions of real world data and make use of prior knowledge.
Many of these requirements involve a tradeoff between accuracy and efficiency, or between accuracy and modeling complexity. These tradeoffs are valued differently depending on the purpose of an entity resolution task. Therefore, despite attempts at generalizability, the suitability of a given entity resolution approach is often highly application-specific.


Finally, one often wishes to understand the uncertainty of the entity resolution process. Errors can occur at the blocking stage due to invalid assumptions or because of a lack of discriminatory attribute information.
Statistically valid quantification of uncertainty requires accounting for possible errors at all stages and propagating uncertainty throughout. This is important when deduplication, record linkage, or entity resolution is used as only the first step of an analysis, as discussed in Section \ref{sec:canonicalization}. Omitting to account for uncertainty can lead to overconfidence and invalid inferences in subsequent analyses.

\subsection{Outline}
We have already described the motivational applications for entity resolution, terminology, and challenges. The remainder of the article focuses on the main types of entity resolution that have been developed and are in use today. In Section~\ref{sec:pipeline}, we provide an overview of the entity resolution pipeline as it is typically considered in the literature. In Section~\ref{sec:determin}, we introduce deterministic and rule-based approaches, which are very popular due to their simplicity, interpretability, and scalability. In Section~\ref{sec:probabilistic} we introduce probabilistic record linkage methods that start with the seminal work in the 1940's -- 1960's, which has led to many advancements and extensions. Next, in Section~\ref{sec:modern}, we review advancements of modern probabilistic record linkage, which include extensions to the seminal Fellegi and Sunter method, Bayesian extensions, and Semi-Supervised and Fully Supervised Record Linkage methods. In Section~\ref{sec:clustering}, we review clustering-based approaches to entity resolution. We cover clustering tasks that are specifically post-processing steps, graphical entity resolution methods, and microclustering models. Finally, in Section \ref{sec:canonicalization}, we review canonicalization and data fusion methods. We conclude in Section~\ref{sec:discussion} with a discussion of open research problems, and provide resources of open source software and data sets in the supplementary materials.

\section{The Entity Resolution Pipeline}
\label{sec:pipeline}

Entity resolution is commonly presented as a pipeline comprising four main stages \citep{christen_data_2012, dong_big_2015} as represented below:  
\begin{equation*}
    \textit{attribute alignment} \rightarrow \textit{blocking} \rightarrow \textit{record linkage} \rightarrow \textit{canonicalization}.
\end{equation*}

In the first stage, \emph{attribute} or \emph{schema alignment}, records are parsed as to identify a set of common attributes among the datasets. For example, often records are stored as a long string of information, such as  ``John Smith lives at 23 Main St W. in Durham, N.C." In order to apply record linkage methods, this information must be broken up into a common set of attributes. First, words and punctuation such as ``lives", ``at", ``in", and ``." are removed. Next, the record is broken up to contain the following standardized attributes:  \textit{civic number}: ``123'', \textit{full road name}: ``Main Street West'', \textit{municipality}: ``Durham'', and \textit{state}: ``NC.''



In the second stage, \textit{blocking}, similar records are grouped into \textit{blocks}. Only records appearing in the same block will then be compared; records that do not appear in the same block are automatically determined to be non-matches.
The simplest blocking method is known as \emph{traditional blocking}, which chooses certain fields (e.g., gender and year of birth) and places records in the same block if and only if they agree on all fields. This amounts to a deterministic a priori judgment that these fields are error-free. Other \emph{probabilistic blocking} methods use probability, likelihood functions, or scoring functions to place records in similar partitions. For example, techniques based upon locality sensitive hashing (LSH) utilize all the features of a record, and they can be adjusted to ensure that all the blocks are manageably small \citep{steorts_comparison_2014}.

\cite{Christophides2019} provides an in-depth review of blocking methods. \cite{steorts_comparison_2014} reviews many types of blocking methods in the literature and provides comparisons. \cite{murray2016probabilistic} reviews  deterministic and probabilistic methods and illustrates how blocking methods integrate with probabilistic record linkage. Blocking is not the main emphasis of the review paper, so we recommend these articles as well as \cite{christen_data_2012} for thorough reviews of the blocking literature. 

In the third stage, \emph{entity resolution} or \emph{record linkage}, duplicate records (or \textit{coreferent} records which refer to the same entity) are identified. In this article, we review deterministic, rule-based approaches (Section \ref{sec:determin}), probabilistic record linkage and modern extensions (Sections \ref{sec:probabilistic} and \ref{sec:modern}), and clustering-based approaches to entity resolution (Section \ref{sec:clustering}).
Finally, in the fourth stage, \emph{merging}, \emph{data fusion}, or \emph{canonicalization}, entities resolved as matches in the third stage are merged to produce a single representative record. We discuss this in Section \ref{sec:canonicalization}.

\section{Deterministic Approaches}
\label{sec:determin}

In this section, we review simple rule-based approaches to entity resolution, and we provide an example from a case study in El Salvador to motivate the need for probabilistic record linkage.

\paragraph{Rules and Scoring Functions.}
In practice, the most commonly used record linkage methods are based on a series of deterministic rules involving the comparison of record attributes.
A simple example is \textit{exact matching}, where two record pairs are linked if they agree on all common attributes. An extension, \textit{off by k-matching,} states that two record pairs are a match if they match on all common attributes except $k$, where $k$ is an integer larger than 0. This method is commonly used when all the attributes are categorical as it tends to perform well, as opposed to when textual variables are introduced.

Record attributes are often distorted by noise (due to data entry errors, variant spellings, outdated records, etc.). Naturally, linkage rules should account for slight differences between attributes. One simple way to quantify such differences for names, addresses, and other textual attributes is via string distance functions. For westernized words, string distance functions such as the \textit{edit} distance (or Levenshtein distance) are used to account for deletions, intersections, and substitutions. The Jaro-Winkler distance works well for the comparison of strings with fewer than 9 characters. For more information regarding distance functions, we refer the reader to \cite{cohen2003comparison}.
Given such measures of similarity between attributes, thresholds can be introduced as to determine matching attributes.

More complex systems involve conjunctions and disjunctions of multiple rules. For example, we may wish to match two records if gender and city of birth agree, which is an example of a conjunction.
More precisely, a \emph{conjunction} of two rules means that the two rules must be simultaneously satisfied for records to be matched. A \emph{disjunction} states that either one of two rules must be satisfied (rule $A$ \textbf{or} rule $B$ is satisfied). Combining these logical operations over a set of base attribute comparisons allows the construction of sophisticated rule-based systems such as a \emph{disjunction of conjunctions}. An example of this rule would be to match records if \textbf{(gender and city of birth agree) OR (date of birth year and city of birth agree)} are satisfied.

In practice, these rule-based systems are carefully crafted for the application at hand. We refer the reader to \citep{Potosky1993} for an example of such a system built for linking the Surveillance, Epidemiology, and End Results (SEER) Program tumor registry with Medicare claims information. In the context of human rights applications, the Human Rights Data Analysis Group (HRDAG) also uses rule-based systems as part of the blocking stage of their entity resolution pipeline. In a blog post, Patrick Ball describes how training data is used to algorithmically determine a disjunction of conjunctions of specified deterministic rules \citep{Ball2016}.

\paragraph{An Illustrative Example from El Salvador.} 

To more fully understand the applicability of rule-based methods for entity resolution, we return to the case study of El Salvador introduced in Section \ref{sec:introduction}.  After the peace agreement of 1992, the United Nations Commission on the Truth (UNTC) for El Salvador invited members of Salvadoran society to report war-related human rights violations \citep{betancur1993madness}. Testimonials were provided many years after the civil war, resulting in possible distortions in the data. In addition, there is no ground truth for this data set, except for two departments which were hand-matched by \cite{sadinle_detecting_2014}. In total, the \texttt{UNTC} data set contains 5395 records. Attribute information available is full name, full date of death, municipality, and department.

Due to the amount of noise that is thought to be in this particular data set, it is often impossible to be certain that two records match or do not match. For instance, consider a small subset from the UNTC data set in Table \ref{table:sv}. Records 4 and 5 can be thought of referring to the same individual, but they could also be father and son. For records 1 --- 3, there are errors in the month and day of death. Exact matching is not applicable here given these errors. The best that can be done, in many cases, is to estimate the probability that records match. \cite{sadinle_detecting_2014} therefore proposed a probabilistic record linkage model where deterministic rules were only used as a blocking criteria. Probabilistic record linkage is introduced next in Section~\ref{sec:probabilistic} and his particular approach is discussed in Section~\ref{sec:BayesianFS}.

\begin{table}[h]
\begin{center}
\begin{tabular}{ccccccc}
\toprule
Record & Given name & Family name & Year & Month & Day & Municipality\\
\midrule
1. &JOSE & FLORES & 1981 & 1 & 29 & A \\
2. &JOSE & FLORES & 1981 & 2 & NA & A \\
3. &JOSE & FLORES & 1981 & 3 & 20 & A \\
4. &JULIAN ANDRES & RAMOS ROJAS & 1986 & 8 & 5 & B\\
5. & JILIAM  & RMAOS  & 1986 & 8 & 5 & B\\
\bottomrule
\end{tabular}
\end{center}
\caption{Illustrative example of duplicated records in the UNTC data set reproduced from Table~1 of \cite{sadinle_detecting_2014}. 
Note that record 5 most likely has errors,  where ``RMAOS" should be ``RAMOS,'' due to the processing of list photocopies using Optical Character Recognition. These errors were corrected in \cite{sadinle_detecting_2014}.}
\label{table:sv}
\end{table}%

While deterministic approaches are appealing for their simplicity and computational scalability, they should be used to help develop intuition, as baseline comparison methods, or as part of a blocking stage. Empirical studies comparing deterministic and probabilistic record linkage techniques used for epidemiological research have shown consistent improvements of probabilistic methods over deterministic approaches \citep{Dusetzina2014, Gomatam2002, Campbell2008, Tromp2011, Avoundjian2020}. Other empirical studies in statistics and computer science have illustrated the superiority of probabilistic approaches over deterministic ones for a case study in Syria \citep{Sadosky2015, chen2018unique}. Other literature has surveyed probabilistic versus deterministic methods more broadly in terms of comparisons, also finding more merit in probabilistic methods \cite{steorts_comparison_2014, murray2016probabilistic}.

\section{Probabilistic Record Linkage}
\label{sec:probabilistic}

Before discussing modern probabilistic record linkage (Section \ref{sec:modern}), we {first take a step back to} introduce the earliest published work on record linkage (Section~\ref{sec:Dunn}). Section~\ref{sec:FS} {then} introduces the Fellegi-Sunter framework, its probabilistic interpretation, and its underlying assumptions. 

\subsection{Dunn's ``Book of Life'' and Early References}\label{sec:Dunn}

The first known paper on record linkage was by 
\cite{Dunn1946}, who defined record linkage as the process of assembling  pieces of information that refer to the same individual. 
For example, birth, marriage, health, and death records are spread across different time periods and locations. Assembling this information for death certificates, identity certifications, and official statistical purposes was and still is a motivating application. Thus, in 1946, Dunn wrote: 

\begin{quote}
``Each person in the world creates a Book of Life. This Book starts with birth and ends with death. Its pages are made up of the records of the principal events in life. Record linkage is the name given to the process of assembling the pages of this Book into a volume.''
\end{quote}

Interestingly, \cite{Dunn1946} framed record linkage as an entirely logistical problem rather than an algorithmic one. He argued that the widespread use of the birth certificate number together with a centralized index would effectively bind together all relevant individual records into their ``Book of Life." This would facilitate and reduce the cost of administrative processes. This has been partly realized with the use of social security numbers, passport numbers, and other unique identifiers.

However, it is often the case that a unique identifier
(centralized index) is not available, especially when there are multiple databases. In addition, a reliable unique identifier such as a social security number may not be available for a variety of reasons. For example, such identifiers often cannot be shared across all databases, are not available due to privacy reasons, or may not exist altogether.
In such situations, record linkage then becomes both a methodological and algorithmic problem --- what can best be used to identify records which refer to the same individual, given noisy, uncertain information?

\cite{newcombe_automatic_1959} provided the first automatic computer-based solution. In their application, the authors wished to link 34,138 birth records from 1955 British Columbia to 114,471 records of marriage from 1945. The aggregated information would then be used in demographic studies \citep{Newcombe1962b, Newcombe1965, Newcombe1965b, Newcombe1969}. There was no unique identifier for the couples, but the birth and marriage records shared full family names, first initials and birth places, ages on some records, and location of the child birth or marriage events. No single piece of information was entirely reliable, but together they could be used for more accurate record linkage.

Their method consisted of two steps. First, they used blocking to reduce the number of comparisons between pairs of records. Specifically, to account for variations in spelling, records were blocked (indexed) based on the Soundex coding of the names.\footnote{The Soundex coding scheme was introduced by Margaret K. Odell and Robert C. Russell (see U.S. patents 1261167 (1918) and 1435663 (1922)). It codifies names by the first letter and by a string of three numbers, with the property that phonetically similar names often share the same code.} Table~\ref{tab:my_label} provides an example of attribute information from compared marriage and birth records from the original \cite{newcombe_automatic_1959}. Second, when Soundex coding agreed between two records, they computed a likelihood ratio comparing the hypothesis that the record pair were a match to the hypothesis that they were not.
If this likelihood ratio exceeded a threshold, then the two records were linked (declared co-referent); otherwise, they were not linked (declared non co-referent).\footnote{The likelihood ratio of \cite{newcombe_automatic_1959} seems to correspond to the computations in ``Method~1'' of \cite{fellegi_theory_1969}, although little information is given by \cite{newcombe_automatic_1959} about the specifics of their procedure.} Studies of the accuracy of the linkage showed about $98.3\%$ of the true matches were detected, and about $0.7\%$ of the linked records were not actual matches. In terms of computational speed, 10 records could be linked every minute on the Datatron 205 computer.

\begin{table*}[h]
    \centering
    \begin{tabular}{rcc}
    \toprule
    Attribute Information & Marriage record & Birth record\\
    \midrule
        Husband's Soundex name code & A300 & A300\\
        Wife's Soundex name code & B600 & B600\\
        Husband's family name & \textbf{Ayad} & \textbf{Ayot} \\
        Wife's family name & Barr & Barr\\
        Husband's initials & J Z & J Z\\
        Wife's initials & \textbf{M} T & \textbf{B} T\\
        Husband's birth province & AB & AB\\
        Wife's birth province & PE & PE\\
        \bottomrule
    \end{tabular}
    \caption{Example of attribute information from marriage and birth records. This table is adapted from Table I of \cite{Newcombe1969} and translated from French to English. AB and PE represent the Canadian provinces of Alberta and Prince Edward Island. Only the initials of the first and middle names are provided in this data.
    }
    \label{tab:my_label}
\end{table*}

In short, the work of  \cite{newcombe_automatic_1959} and \cite{Newcombe1962} introduced key ideas for record linkage in an application to demographic data, where blocking (indexing) was used to make the problem computationally tractable. They proposed an informal statistical approach based on a likelihood ratio test, where the pipeline was fully automated and required no training data.

\subsection{A Theory of Record Linkage}\label{sec:FS}

We now turn to the most widely utilized record linkage task --- the Fellegi-Sunter method, where \cite{fellegi_theory_1969} formalized the approach of \cite{newcombe_automatic_1959} in a decision-theoretic framework. For a given pair of records, three possible actions are considered: to \textit{link}, to \textit{possibly link}, or to \textit{not link}. The goal is to minimize the number of \textit{possible links}, while controlling for two types of conditional error probabilities: (1) the probability $\mu$ of linking when the records do not match
, and (2) the probability $\lambda$ of not linking when the records do match. In the framework of \cite{fellegi_theory_1969}, an optimal linkage procedure at the levels $\mu$ and $\lambda$ is one which attains these error rates while minimizing the number of \textit{possible link} assignments. A ``fundamental theorem for record linkage'' demonstrated by the authors shows that the optimal linkage procedure corresponds to a likelihood ratio test.

Consider two records $a$ and $b$, and let the \textit{comparison vector} or an \textit{agreement pattern} $\gamma$ represent the level of agreement/disagreement between the two. 
For instance, $\gamma$ could be
$$\text{``initials between the records agree and are J\&M"}$$
In practice, $\gamma$ is usually decomposed as $\gamma = (\gamma_1, \gamma_2, \dots, \gamma_k)$, where each $\gamma_i$ corresponds to a comparison between a particular attribute (name, age, etc.) of the record pair. 
One can consider binary comparisons, where $\gamma_i \in \{0,1\}$ represents agreement or disagreement between record attributes, as well as more detailed comparisons involving the specific value for which there is an agreement (such as $\gamma_i = \text{``initials agree and are J\&M''}$). 
While the choice of comparisons to consider is theoretically arbitrary, it involves practical considerations which are discussed in the following paragraphs. 

Let $m(\gamma)$ be the probability of observing the comparison vector $\gamma$ for two records that are an actual match, let $u(\gamma)$ be the probability of observing the comparison vector $\gamma$ for two records that are not a match, and let $W(\gamma) = \log m(\gamma) - \log u(\gamma)$ be the log likelihood ratio (the \textit{matching weight}). {A large value of $W(\gamma)$ is indicative of a match, while a small value is indicative of a non-match.}
The optimal linkage procedure, with thresholds $T_\mu$ and $T_\lambda$ corresponding to the error rates $\mu$ and $\lambda$  is given as follows:\footnote{See Section 3.7 of \cite{fellegi_theory_1969} for the definitions of $T_\mu$ and $T_\lambda.$ These threshold are selected such that the error rates are less than both $\mu$ and $\lambda.$ In addition, there is no other rule that can produce a smaller clerical review area that controls the error rates at this level.}

Assume a pair of records with a comparison vector $\gamma$ and matching weight $W(\gamma).$ 
\begin{itemize}
\item We link the pair of records if $W(\gamma) > T_\mu$
\item We call the pair of records a possible link if $T_\mu \geq W(\gamma) > T_\lambda$
\item We do not link the pair of records if $T_\lambda \geq W(\gamma).$
\end{itemize}
This optimal linkage procedure ignores the boundary cases, however, these details are in  Appendix~1 of \cite{fellegi_theory_1969}.

Two methods are proposed by \cite{fellegi_theory_1969} to estimate the $m$ and $u$ probabilities. In both methods below, the author's assumed conditional independence between the attribute comparisons $\{\gamma_i\}_{i=1}^k$ given the true underlying match/non-match status of the record pairs.

First, the authors considered 
 \emph{detailed comparison} vectors, which provide both an indication of agreement or disagreement for each attribute and a precise shared value in the case of an agreement. This allows one to exploit specific information about the record's attributes. For instance, two records agreeing on the less common name ``Xander'' are more likely to be a match than two records which only agree on the first name ``John.'' In applications, it is often helpful to exploit such frequency information.
 Under this assumption, the authors used the frequency distribution of the record's attributes, together with prior information about error rates, to obtain estimates of the $m$ and $u$ probability distributions.

Second, the authors considered \emph{binary comparisons}, where each $\gamma_i$ is a binary variable indicating agreement or disagreement the records' $i$th attribute. The distributions $m$ and $u$ can then be estimated from the observed frequencies of agreement or disagreement between these fields. In particular, they derived analytical formulas to estimate $m$ and $u$ when only three fields are under comparison. 

\cite{Winkler1988} extended the above estimation methods by proposing the use of  the EM algorithm to estimate the $m$ and $u$ distributions both in the context of detailed comparisons between fields (where particular agreement values are also taken into consideration) and  binary comparisons. Independently, \cite{Jaro1989} proposed the EM algorithm for binary comparisons and considered its application for matching the 1985 test census (dress rehearsal)
of Tampa, Florida, to an independent post-enumeration survey as to evaluate the census coverage.

\paragraph{Interpretation of the Probability Model.}

While the Fellegi-Sunter approach was introduced in a decision-theoretic framework, it can more easily be interpreted through its underlying probability model and using posterior match probabilities.

In order to explain this, recall that probabilistic record linkage has two main steps. First, one compares record pairs to obtain comparison data. Second, one uses this comparison data to classify record pairs as being links, possible links, or non-links. The second step relies on a probability model for the comparison data. That is, the comparison vectors $\gamma$ are assumed to be independently distributed using the following mixture model:
$$
    p(\gamma) = \lambda m(\gamma) + (1-\lambda) u(\gamma),
$$
where $\lambda > 0$ is the probability that a randomly chosen comparison vector corresponds to a matching pair of records. The methods proposed by Fellegi-Sunter, as well as the EM algorithm proposed in \cite{Jaro1989, Winkler1988}, provide estimates $\hat \lambda$, $\hat m$ and $\hat u$ of the parameters $\lambda$, $m$, and $u$. 

Denote a true match by $M.$
Using Bayes rule, one can express the probability that two records match given {their} comparison vector $\gamma$, as
\begin{equation}\label{eq:match_posterior_prob}
    p(M \mid \gamma) = {\lambda m(\gamma)}/{p(\gamma)} =1 - \left(1 + \frac{m(\gamma)}{u(\gamma)} \frac{\lambda}{1-\lambda}\right)^{-1}.
\end{equation}
The left-hand side of equation \ref{eq:match_posterior_prob} is the posterior probability of a match, and the right-hand side shows how it can be obtained as a monotonous transformation of the likelihood ratio $m(\gamma)/u(\gamma)$ considered by \cite{fellegi_theory_1969}. Therefore, as noted in \cite{larsen_2001}, thresholding the posterior probability to assign links is equivalent to using a likelihood ratio test and the optimality result of \cite{fellegi_theory_1969} also applies in this context.

\cite{Tepping1968} considered directly using the posterior probabilities in equation~\ref{eq:match_posterior_prob} to minimize the \textit{expected cost} of linkage procedures. For example, suppose there are three actions{:} to link (denoted $A_1$), to specify a possible link ($A_2$), or not to link ($A_3$).
Let U denote the records are not a match. Then we associate 
 each action $A_i, i=1,2,3$ with costs $C(A_i; M)$ and $C(A_i; U).$ 
The expected cost of action $A_i$ given $\gamma$ is 
$$
    C(A_i \mid \gamma) = C(A_i; M) p(M \mid \gamma) + C(A_i; U) (1-p(M \mid \gamma)).
$$
Given a comparison vector $\gamma$, the optimal decision is then to take the action $A_i$ with smallest expected cost. Since the expected costs $C(A_i \mid \gamma)$ are linear in the posterior probability of a match $p(M \mid \gamma)$, this optimal decision procedure is also equivalent to a likelihood ratio test with adjusted thresholds.
While estimates of the posterior probabilities $p(M \mid \gamma)$ can be obtained in the Fellegi-Sunter framework, \cite{Tepping1968} suggested to directly estimate them through sampling and clerical review. For example, consider four fields: name, surname, age and birth city. Using binary comparisons between these fields, there are $2^4$ possibilities for the comparison vector $\gamma \in \{0,1\}^4,$ where each defines a comparison class. Under the approach of \cite{Tepping1968}, one samples record pairs within each class and uses these to estimate the probability $p(M \mid \gamma)$ directly. Next, one can determine the contribution of this comparison class to the total expected cost which is obtained by summing the cost of the optimal decision over all record pairs. Finally, the classes with the largest associated cost can then be further subdivided as to improve the quality of the linkage.

\cite{DuBois1969} considered a particular case of Tepping's Bayesian decision framework and proposed estimating the $m$, $u$ and $\lambda$ parameters using training data (manually classified record pairs) and maximum likelihood, while allowing for missing values. Tepping's framework was also revisited by \cite{Verykios2003} who also considered blocking and an empirical analysis.

\paragraph{Assumptions of Fellegi-Sunter.}
The Fellegi-Sunter framework relies on crucial simplifying assumptions which are not expected to hold in practice. 

The first assumption is that comparison vectors between the records pairs should be independent from one another.
{This is usually not satisfied in practice.}
For example, when \cite{newcombe_automatic_1959} linked birth and marriage records, it was known that two different marriages could not result in the same birth. This constraint induces dependencies between comparison vectors, and applying the Fellegi-Sunter procedure can lead to impossible linkage configurations when this is not taken into consideration. Moreover, the assumption is not satisfied in the context of one-to-one linkage, where each record in a first file can be linked to at most one other record in a second file. The issue was faced by \cite{Jaro1989}, where each census record could be matched with at most once record in the post-enumeration survey. To account for this, the author proposed to precede the Fellegi-Sunter likelihood ratio test by a blocking stage, which established a one-to-one constraint. That is, he constructed a first set of one-to-one candidate links between the census and the post-enumeration survey. The candidate links were chosen as to maximize the sum of their matching weights, while satisfying the one-to-one constraint. Among these candidate pairs, the Fellegi-Sunter likelihood ratio test (thresholding the matching weight) was then used to establish the link, possible link or non-link assignment. The same kind of issue also arises when enforcing transitivity between links. In many cases, knowledge that records $A$ and $B$ are a match, and that $B$ and $C$ are a match, means that necessarily $A$ and $C$ are also a match. This introduces another form of dependency between records and their comparison. Methods to formally address issues of this kind in the Fellegi-Sunter framework can be complex and computationally demanding, especially when more than two databases are considered \citep{Sadinle2013}.

The second assumption is that the $m$ and $u$ distributions are known or can be adequately estimated. And to be practically feasible, their estimation relies on simplifying assumptions which usually do not hold.
For one thing, the estimation methods discussed so far require conditional independence between the comparison of different record attributes, given the true match/non-match status of the record pairs. \cite{Smith1975} first remarked that this conditional independence assumption may not hold in practice. \cite{Thibaudeau1993} (see also \cite{Armstrong1992, Winkler1992, Winkler1993}) proposed log-linear models with interaction terms to account for dependencies between field comparisons and showed improved performance in some applications. {\cite{Xu2019} have illustrated recently that the conditional independence assumption can work quite well in practice depending on the discriminatory power of the linkage variables. The authors applied latent class models with a conditional dependence structure informed by the true match status of manually reviewed record pairs. In one scenario, where the attributes have poor discriminating power, the conditional dependence models yields improved matching accuracy compared to the FS model. In a second scenario, where the attributes have good discriminating power for linkage, incorporating conditional dependence results in comparable matching accuracy relative to the FS model. This provides guidance to researchers empirically when conditional independence is reasonable and unreasonable. \cite{Daggy2014} also reviewed and evaluated the use of conditional dependency models for record linkage applications.}

Given that the assumptions of Fellegi-Sunter are often not satisfied, this has led to many extensions in the literature, which we review in section~\ref{sec:modern}.

\section{Modern Probabilistic Record Linkage}
\label{sec:modern}
In this section, we review modern probabilistic record linkage, which includes extensions to the Fellegi-Sunter framework, Bayesian variants of Fellegi-Sunter, as well as semi-supervised and fully supervised classification approaches.

\subsection{Extensions of Fellegi-Sunter}

\label{sec:FS_extensions}

In many applications, neither the procedure of Fellegi-Sunter nor of Tepping is used to set classification thresholds. According to \cite{belin_1995}, for the matching of the 1990 Census with the post-enumeration survey, thresholds were set ``by `eyeballing' lists of pairs of records brought together as candidate matches.'' Part of the reason is that the error rates fixed in the Fellegi-Sunter framework, as well as the false-match rates estimated using equation~\ref{eq:match_posterior_prob}, are not attained in practice \citep{Winkler1992, belin1990proposed, Armstrong1992, belin_1995}. This is due to the various simplifying assumptions and estimation errors involved in the application of such models. Therefore, methods using training data (classified record pairs) have  been proposed to automate and improve the choice of tuning parameters in probabilistic record linkage.

For instance, \cite{belin_1995} proposed to calibrate thresholds and error rates by using training data to fit a mixture model to the \textit{matching weight} distribution. This allowed the authors to quantify uncertainty about the linkage's error rates and to calibrate the Fellegi-Sunter thresholds. \cite{Nigam2000} showed how training data could be combined with unlabeled data as to improve the estimation of the $m$ and $u$ distributions using the EM algorithm {for text classification}. In this semi-supervised framework, \cite{winkler_2000, winkler_2002} and \cite{larsen_2001} considered fitting more complex models allowing for dependencies between field comparisons. 

Recently, \cite{enamorado_using_2019} addressed the problem of scaling record linkage to large data sets by developing an efficient implementation, called \texttt{fastLink}, of the seminal record linkage model originally proposed by \cite{fellegi_theory_1969}. In addition, the authors extended work of \cite{lahiri_2005} in order to incorporate auxiliary information such as population name frequency and migration rates into the merge procedure and conduct post-merge analyses, while accounting for the uncertainty about the merge process. The authors used parallelization and efficient data representations in order to merge millions of records in near real time on a laptop computer, and provide an open source R package for their proposed methodology.

\subsection{Bayesian Fellegi-Sunter}\label{sec:BayesianFS}

\cite{Fortinietal01} proposed the first Bayesian approach to entity resolution in the specific case of record linkage between two databases, with no duplication within each database (bipartite record linkage).\footnote{For a detailed explanation of bipartite record linkage, see \cite{sadinle_bayesian_2017}.}
Their approach can be interpreted as a Bayesian version of the Fellegi-Sunter framework (Bayesian FS). A prior on the number of matching pairs is considered, together with a prior on the matching configuration matrix
\footnote{The matching configuration matrix, or coreference matrix, indicates the linkage structure between two databases. 
If we denote by $i$ a record in the first database and $j$ a record in the second database, then this matrix has entries $c_{i,j} \in \{0,1\}$, with $c_{i,j} = 1$ if records $i$ and $j$ are linked and $c_{i,j} = 0$ otherwise.}
and a Dirichlet prior on the $m$ and $u$ distributions. These parameters are then estimated through Markov Chain Monte Carlo. The Dirichlet distribution is a flexible prior which allows dependencies between the components of comparison vectors, and the Bayesian formulation allows to easily incorporate constraints on the linkage. More specifically, assuming no duplications within a database naturally implies that a record from one database can be linked with at most one record from the other database (this is often referred to as a maximum one-to-one linkage assignment). This constraint is easily incorporated in the Bayesian model through the prior on the matching configuration matrix \citep{Larsen05}.

More recently, \cite{sadinle_detecting_2014} has proposed a Bayesian FS method for deduplication, where the author also relies on comparison vector data. He also used a likelihood ratio similar in spirit to \cite{Fortinietal01} and \cite{fellegi_theory_1969}. Sadinle's main innovation is the consideration of a prior on the matching configuration matrix which imposes \textit{transitive closures} ---records are partitioned into groups which are thought to refer to the same entity. In contrast with ad hoc approaches to resolving intransitivity, this approach allows quantification of uncertainty about the partition of records through a posterior distribution. In later work, \cite{sadinle_bayesian_2017} extended the above framework for bipartite record linkage. In addition, the authors derive Bayes estimates under a general class of loss functions, which provides an alternative to the Fellegi-Sunter decision rule. Both the work of \cite{sadinle_detecting_2014} and \cite{sadinle_bayesian_2017} apply their proposed methodology with deterministic blocking rules to a case study on human rights in El Salvador, which was a motivating example in Section~\ref{sec:introduction}. In addition to proposing new methodology, \cite{sadinle_detecting_2014} performed hand-matching on a small set of the dataset such that pairwise evaluation metrics could be utilized.  In a recent extension of both these methods, \cite{sadinle2018bayesian} proposed a two-stage approach to record linkage and multiple systems estimation (MSE), where the author first removes duplications from the El Salvador dataset and then utilizes a MSE methodology in order estimate the unknown population size.

One difficulty facing Bayesian FS is their computational burden. In other recent work, \cite{mcveigh2019scaling}  considered this issue by proposing a blocking approach based on simpler probabilistic record linkage techniques. That is, the output of more simple non-Bayesian probabilistic record linkage is used to perform ``post-hoc blocking,'' after which a Bayesian FS method is used for coherent modeling and uncertainty quantification. This allows the authors to scale their proposed method to voter registration and census data sets with million of entries.

\subsection{Semi- and Fully Supervised Classification Approaches}

The approaches of \cite{belin_1995, Nigam2000, winkler_2002, winkler_2000} and \cite{larsen_2001} discussed in Section \ref{sec:FS_extensions} were \textit{semi-supervised} \citep{Chapelle2006}. Semi-supervised methods use a relatively small amount of manually classified record pairs, known as labeled pairs, to improve upon unsupervised probabilistic record linkage. In this section, we review other semi-supervised methods. In addition, we review fully supervised methods, which focus on classifying record pairs as a first step to entity resolution. 
To summarize, all of the methods reviewed in this section are used to obtain predicted match probabilities for record pairs.

\paragraph{Semi-Supervised Approaches.}

Following the terminology of \cite{Chapelle2006}, \textit{generative} semi-supervised approaches target the joint likelihood of the labeled and unlabeled data as in \cite{Nigam2000} and \cite{larsen_2001}. Building on this framework, \cite{Enamorado2019} proposed an active learning algorithm which iteratively requests labels for specific record pairs. Other active learning approaches are proposed in \cite{Sarawagi2002, Bellare2012, Wang2015} and \cite{Christen2016}.
\textit{Change of representation} semi-supervised approaches use unsupervised learning as a first step to summarize the data (such as performing dimensionality reduction), before using a supervised algorithm for further analysis. For instance, \cite{belin_1995} used the unsupervised Fellegi-Sunter framework to obtain univariate matching weights for all record pairs, before using labeled examples to fit a mixture model to the matching weights. This allows the authors to calibrate the model and potentially select better thresholds.
\textit{Self-learning} algorithms generalize the semi-supervised EM algorithm considered in \cite{Nigam2000} and \cite{larsen_2001} to model-free classifiers. In this framework, \cite{Kejriwal2015} combined self-learning and boosting of random forests and multi-layer perceptrons as to obtain good performances on ER tasks using only small amounts of labeled pairs.

\paragraph{Fully Supervised Approaches.} \label{sec:semi_supervised}
Fully supervised methods do not exploit information provided by unlabeled examples; instead they rely on larger amounts of labeled pairs. Given the significant class imbalance when considering record pairs (very few pairs match), vast amounts of reliable training data or carefully selected training data is required for the use of these methods.
This training data may come from crowdsourcing \citep{Sarawagi2002, Wang2012, Vesdapunt2014, frisoli2019novel}, from extensive manual record linkage efforts \citep{trajtenberg2008identification, Azoulay2011, Bailey2017}, or it may be automatically generated using unsupervised methods as to obtain an \textit{approximate} training set \citep{Torvik2005, Christen2007, Christen2008}. In practice, the amount of reliable training data necessary to train sophisticated learning algorithms such as deep neural networks \citep{Gottapu2016, Ebraheem2017, Mudgal2018, Kooli2018, Kasai2020} is rarely available for ER tasks\footnote{For instance, \cite{Kooli2018} uses over 10 million examples of labeled record pairs (corresponding to more than 3,000 resolved individual records) in an application in order to train deep neural networks. More recently, \cite{Kasai2020} considered the issue of training deep neural networks for entity resolution with fewer labels, using active and transfer learning.}
and simpler classifiers (such as logistic regression, decision trees, random forests,
Bayesian additive regression trees, and many other \citep{hastie_2001}) are preferred. 

To give an example of how such methods are used in practice, consider the work of \cite{ventura2015seeing}, a case study for inventor disambiguation in the bibliographic database of U.S. Patent and Trademark Office (USPTO) patents. The authors proposed a supervised method based on random forests
%
for deduplication. Their training data was constructed from the curriculum vitae of inventors in the field of optometrics as well as from a previous study on ``superstar'' academics in the life sciences \citep{Azoulay2007, Azoulay2011}. This allowed them to evaluate the performance of previous methods used in this application \citep{Li2014, fleming2007small} and to train their random forest classifier on labeled comparison vectors of record pairs. Afterwards, applying their entity resolution approach to other records in the USPTO database consisted of a four-stage pipeline. First, they use blocking where, in each block, they calculated
comparison vectors for each record pair. Second, the authors calculated the  predicted probability of a match using their random forest classifier applied to these comparison vectors. Third, the predicted probability was converted into an estimate of the dissimilarity between each pair of records. Fourth, the authors utilized  single linkage hierarchical clustering corresponding to the dissimilarity scores in the previous step to enforce transitive closures among record pairs. Clusters were determined by cutting the dendogram (tree) at a threshold. Finally, all the clustering results were combined across blocks to obtain a final set of de-duplicated records. Such clustering approaches to ER, used either directly or as a follow-up to pairwise classification, are discussed next in Section~\ref{sec:clustering}.

\section{Entity Resolution as a Clustering Problem}\label{sec:clustering}

The methods discussed so far focused on estimating the probability of a match between pairs of records given their comparison vector. This pairwise match probability provides a measure of uncertainty about specific links, where the corresponding false match and false non-match rates (or precision and recall) are pairwise evaluation metrics of performance. With the exception of Bayesian FS, these methods treat record pairs as being independent of one another, without accounting for the consequences of transitivity or other constraints on the linkage structure. This limits their applicability, and specifically in the context of linking more than two databases when duplicates are present across and within databases. For instance, while \cite{sadinle_multi_2} and \cite{Sadinle2013} have proposed a principled generalization of the Fellegi-Sunter framework to multiple databases; this is not computationally tractable in the presence of more than three databases.

Much of the literature has therefore advocated for a clustering-based approach to entity resolution and deduplication which can integrate multiple databases \citep{Monge1997, Cohen2002, liseo_2011, sadinle_detecting_2014, Ventura2014, steorts14smered, steorts_entity_2015, Rahm2016, zanella_flexible_2016, sadinle_bayesian_2017, marchant_distributed_2019}.
In this context, the goals shift. Instead of linking record to record, the goal is to cluster records to their true (unknown, latent) entity.

A large portion of this literature uses clustering as a second step to probabilistic record linkage to enforce transitivity of the output \citep{Hassanzadeh2009, Christophides2019}. These approaches, presented in Section \ref{sec:clust_post_process}, are similar to the deterministic entity resolution techniques discussed in Section \ref{sec:determin} since they generally provide no probability statement regarding the resolved entities. Other clustering approaches are \textit{model-based}, and in particular we focus on \textit{graphical entity resolution} in Section \ref{sec:clust_model_based}. By probabilistically modeling the relationship of records to the latent entities to which they refer, these methods naturally provide uncertainty quantification regarding the clustering structure \citep{steorts_entity_2015, steorts_bayesian_2016}.
Finally, entity resolution can be viewed as what we refer to as a \emph{microclustering} problem, meaning that the size of the latent clusters grows sub-linearly as the number of records grows. This means that entity resolution does not experience power law (linear) growth as many traditional clustering tasks. We discuss \emph{microclustering} in Sections \ref{sec:microclustering} and \ref{sec:feasibility}.

\subsection{Clustering as a Post-Processing Step} \label{sec:clust_post_process}

Many clustering approaches to entity resolution are based on pairwise similarities,  pairwise match probabilities, or determined links and non-links. Therefore, these can  be seen as post-processing the result of other pairwise record linkage procedures. They are used to resolve intransitivites in the linkage method and ensure a coherent output. There is a vast literature {on the subject}; we only review a selection of the proposed methodology for entity resolution. We refer the reader to \cite{Hassanzadeh2009, Naumann2010, christen_data_2012, Han2011} and \cite{Christophides2019} for more exhaustive reviews.

One of the first references in this area is \cite{Monge1997}, who framed entity resolution as a clustering problem. Specifically, they proposed that one should detect the connected components in the undirected graph of pairwise links (see also \cite{hernandez_1995, Hernandez1998}). Pairwise links were determined iteratively. At any given step, only records which were not in the same connected component were compared in order to determine the match/non-match status. This is known as a \textit{dynamic connectivity} problem. It allowed the authors to resolve intransitivities in pairwise matching while avoiding superfluous comparisons. The idea of clustering through connected components is computationally efficient and has recently been exploited as part of a blocking stage in \cite{mcveigh2019scaling}.
A more sophisticated technique, \textit{correlation clustering} \citep{Bansal2004}, maximizes the number of links within clusters plus the number of non-links across clusters. This approach was originally introduced in the context of document classification. It does not require the specification of the number of clusters and it is obtained from a single meaningful objective function. However, correlation clustering is NP-hard \citep{Filkov2003, Bansal2004} and in practice variants and approximate solutions are used \citep{Charikar2003, Ailon2008, Gionis2007, Hassanzadeh2009}. 
Another approach is \textit{hierarchical agglomerative clustering} \citep{Johnson1967, hastie_2001}, which \cite{Ventura2014} advocated in conjunction with ensemble classifiers for large scale entity resolution. \cite{ventura2015seeing} applied this method for inventor disambiguation in the USPTO data set as discussed in Section~\ref{sec:semi_supervised}. 

\subsection{Graphical Entity Resolution}\label{sec:clust_model_based}

We now turn to model-based clustering approaches which allow quantification of uncertainty about the clustering structure. \cite{Bhattacharya2006} built on the Latent Dirichlet Allocation (LDA) model to this end, where the goal in their application was to resolve individual authors in bibliographic databases. Their approach leveraged co-authorship groups (analogously to \textit{topics} in LDA) in order to support the entity resolution process. They probabilistically modeled the unkown set of individual authors, the authors' group membership, as well as possible distortions in authors' names. Posterior inference was carried out using Gibbs sampling.

In a similar spirit, \cite{liseo_2011} proposed a new model for record linkage which, instead of linking record to record, linked records to \textit{latent individuals}. The authors used the hit and miss model of \cite{copas_record_1990} as a measurement error model to explain possible distortions in the observed data. This deviates from the Fellegi-Sunter approach as it does not utilize comparison data, instead working with the actual attribute information.

We refer to such approaches, where one recovers a bipartite graph linking records to reconstructed latent entities, as \textit{graphical entity resolution}. More specifically, in \cite{steorts14smered} and \cite{steorts_bayesian_2016}, the authors developed a fully hierarchical-Bayesian approach to entity resolution, using Dirichlet prior distributions over categorical latent attributes and assuming a data distortion model. They derived an efficient hybrid (Metropolis-within-Gibbs) MCMC algorithm for fitting these models, called the Split and MErge REcord linkage and deduplication (SMERED).  
As with other Bayesian approaches, this allows full quantification of uncertainty regarding the number of latent individuals and the clustering structure of records into coreferent groups. In addition, \cite{steorts_bayesian_2016} showed that for the proposed work and the work of \cite{sadinle_detecting_2014} and \cite{tancredi_hierarchical_2011}, the use of a uniform prior on the set of links or non-links, in practice, leads to one having a biased estimation of the sample. This, in turn, led to the development of subjective priors on the linkage structure which have appeared in \citep{zanella_flexible_2016}. In addition to Bayesian models for categorical data, \cite{steorts_entity_2015} extended the above work to both categorical and noisy string data using by proposing a string pseudo-likelihood and an empirically motivated prior called the empirical Bayes (EB) method. The authors provide an R package for this code on CRAN and GitHub called \texttt{blink}. 

Motivated by the computational limitations of \cite{steorts_entity_2015} and a case study of the 2010 Census, \cite{marchant_distributed_2019} have proposed a scalable extension to the \blink\ model for end-to-end Bayesian entity resolution~\citep{steorts_entity_2015}, which they refer to as ``\underline{d}istributed \blink'' or \dblink. Their approach uses probabilistic blocking at the level of the latent entities, which enables distributed inference through a partially collapsed Gibbs sampler while accounting for blocking uncertainty. The authors showed that {\dblink} provides more than a 200 times speed improvement over \blink, allowing the end-to-end Bayesian approach to scale to hundreds of thousand of records.

\subsection{The Microclustering Property} \label{sec:microclustering}

The work of \cite{steorts_entity_2015} and \cite{steorts_bayesian_2016} led to interesting developments both in clustering and in entity resolution. The first is the formalization of the \emph{microclustering property}, which describes the sub-linear growth of clusters in entity resolution (and in other clustering tasks such as community detection). That is, one expects the size of the clusters to grow sub-linearly as the total number of records also grows \citep{zanella_flexible_2016}. Therefore, applying a Bayesian nonparametric (BNP) model which favors large clusters makes little sense in the context where each cluster should correspond to a single true entity. The second development is the proposal of general BNP models which can satisfy the microclustering property. The authors also propose a more scalable algorithm, the chaperones algorithm, which allows for computational speed-ups for entity resolution that are similar in spirit to the Split and Merge approach as also used by \cite{steorts_bayesian_2016}.

\subsection{The Feasibility of Microclustering for Entity Resolution}\label{sec:feasibility}

The aforementioned work has led to considerations regarding the feasibility of entity resolution in the context of microclustering. There are only two papers addressing such implications in the literature to our knowledge. In recent work, \cite{steorts_performance_2017} provided quantitative bounds on the largest number of entities one may hope to resolve, given the number of record attributes, categories within the attributes, noise levels, and a large class of models. Simulations studies are provided that offer guidance for when the bounds are tight and loose in practice that are complementary to a recent paper of \cite{johndrow2017theoretical}. \cite{johndrow2017theoretical} show that unless the number of attributes grows with the number of records, entity resolution is infeasible in certain contexts. This resonates with empirical evidence from the entity resolution literature and is related to the separation between entities going to zero.  The authors suggest that a logarithmic growth in the number of attributes may be sufficient to achieve accurate entity resolution, and if substantiated, this would be a rather encouraging scenario. More work on the curse and blessing of dimensionality in entity resolution is needed to provide further insight on how much information we need to collect for entity resolution problems and provide practical guidance to users \citep{johndrow2017theoretical}.

\section{Canonicalization and the Downstream Task}
\label{sec:canonicalization}

This section reviews the fourth stage of the entity resolution pipeline, which is known as \textit{merging}, \textit{fusion}, or \textit{canonicalization}. The goal here is to merge together records which refer to the same entity as to obtain a single representative record. These may then be used as the basis of further analyses (downstream tasks) such as a regression. Furthermore, there are two main types of methods used to carry out downstream tasks in an entity resolution context. The first is the use of single stage or joint models, where assumptions and data involved in the downstream task can also inform the record linkage process. The second type of method proceeds in two stages, where record linkage is done independently from the downstream task. 

Section~\ref{subsec:canonicalization} discusses methods used to merge records, Section~\ref{subsec:joint} presents joint modeling approaches to entity resolution and downstream tasks, and Section~\ref{subsec:two-stages} finally presents two-stage approaches.

\subsection{Canonicalization}\label{subsec:canonicalization}

Canonicalization, merging, or data fusion is the task of merging groups of records that have been classified as matches into one record that represents the true entity \citep{bleiholder2009data, christen_data_2012}. The earliest proposals of canonicalization were deterministic, rule-based methods, which were application specific and fast to implement \citep{cohen2005incremental}. The existing literature assumes training is available in order to select the canonical record, and authors have proposed optimization and semi-supervised methods to finding the most representative record \citep{yan1999conflict, bohannon2005cost, culotta2007canonicalization}. For a full review of data fusion techniques, we refer to \cite{bleiholder2009data}.

In a recent application motivated by the NCSBE voters data set, \cite{kaplan} provide a unique identifier for voter registration in a principled and reproducible manner. 
In contrast to existing methods for canonicalization, their approach does not rely on training data and can handle categorical, ordinal and numerical attributes that are commonly needed for downstream inference. They instead propose five fully unsupervised methods to choose canonical records from linked data, including a fully Bayesian approach. By performing each stage – entity resolution, canonicalization, and downstream task – in a Bayesian framework, uncertainty is propagated throughout and properly accounted for when reaching final conclusions.

\subsection{Joint Models}\label{subsec:joint}
 
Turning to joint or single-stage modeling approaches to entity resolution and the downstream task, these have been limited to linking two databases and do not easily generalize beyond this framework \citep{DalzellReiter18, steorts2018generalized, gutman_2013, HofRavelliZwinderman17, tang2020}.  Single-stage approaches which jointly model record linkage and the association between key variables, in the context of two files, have been proposed by \cite{HofRavelliZwinderman17} for survival data using a frequentist procedure. On the other hand, \cite{gutman_2013} and \cite{DalzellReiter18} proposed Bayesian methods for regression in a more general framework and motivated by medical applications.
\cite{steorts2018generalized} proposed a joint record linkage and regression model, building off the work of  \cite{tancredi_hierarchical_2011} and \cite{steorts_bayesian_2016}, where they introduce non-parametric priors on the linkage structure. The authors provide evidence in their experimental analysis of when exactly the feedback loop between the task and the record linkage is effective. \cite{tang2020} have proposed a joint model for record linkage and regression, where uncertainty quantification is also propagated. Specifically, the authors extend recent work of \cite{sadinle_bayesian_2017}, focusing on bipartite record linkage, and the authors illustrate that joint modeling can leverage relationships among the dependent and independent variables in the regression to potentially improve the quality of the linkages. In addition, this can increase the accuracy of resulting inferences about the regression parameters. There is a comparison made to the two-stage model of \cite{sadinle_bayesian_2017}, illustrating when improvements can be made. 

We note that, while Bayesian joint models have the advantage of natural error propagation, they can be computationally costly. Moreover, they require knowledge of the model specification up front: if an additional downstream task is required (after the original joint model has been fitted), then the linkage would need to be repeated in a new joint model for valid inference. Because record linkage is the most computationally costly part of joint models, this is a a particularly negative aspect of such single-stage approaches.

\subsection{Two-Stage Models}\label{subsec:two-stages}
Two-stage approaches first perform entity resolution (and canonicalization) before using this as the input of the downstream task. In a Bayesian framework, error regarding the record linkage can still be easily propagated. This is exemplified by the concept of ``linkage-averaging'' discussed in \cite{sadinle2018bayesian} for population size estimation. In contrast with joint models, however, information cannot flow the other way: data or assumptions relevant to the downstream task cannot be used to support the entity resolution process. 

In the context of regression analyses, most of the two-stage literature (record linkage and regression) joins only two databases and often assume that the error from the record linkage task occurs only in the response variable \citep{lahiri_2005, kim12, goldstein12, hof12, tang2020}. For example, \citet{lahiri_2005} addressed the problem of linking two databases under the assumption that they represent a permutation of the same set of records and the linkage error only involves the response variable. They proposed an unbiased estimator (LL) for linear regression, conditional on the matching probabilities provided by the linkage process. \citet{hof12} extended \citet{lahiri_2005} to handle more realistic linkage scenarios under a logistic regression framework. Generalizations of the LL estimator can be found in \citet{kim12}, where estimating equations provide consistent estimators of population quantities. \citet{goldstein12} relaxed these assumptions and considered the matching probabilities as prior information to be used within a multiple imputation scenario.

\section{Discussion}
\label{sec:discussion}


In this article, we have introduced the entity resolution problem as it relates to four important social science issues -- the decennial census, human rights violations, voter registration, and inventor and author {disambiguation}.
Applications are more widespread, {dealing with} medical, housing, and financial databases, among others. We have introduced the main terminology used in the literature, and we have provided the major challenges that researchers face within an entity resolution framework. 
We have presented the pipeline approach (section~\ref{sec:pipeline}), where the four stages consist of attribute alignment, blocking, entity resolution, and canonicalization. This article has focused mainly on entity resolution, as it is the most complex part of the pipeline. Section~\ref{sec:determin} reviewed deterministic rule-based methods that are commonly used in the literature, such as exact matching and scoring rules. Section~\ref{sec:probabilistic} reviewed seminal probabilistic record linkage methods, such as those proposed by Dunn, Newcombe, and Fellegi and Sunter, which led to many modern day extensions (section~\ref{sec:modern}). These extensions can be viewed as extensions of Fellegi and Sunter (frequentist and Bayesian) or as semi- or fully supervised classification approaches. Section \ref{sec:clustering} reviewed entity resolution methods that can be viewed as clustering tasks. These include methods where clustering is a post-processing step, graphical entity resolution, and microclustering models. Section~\ref{sec:canonicalization} reviewed canonicalization, which is the fourth part of the pipeline. This section also reviews the literature that combines entity resolution with downstream tasks, such as  linear regression. 



In the remainder of our discussion, we highlight a few remaining topics which are the subject of active research and which have important practical implications.
First, we discuss the need to rigorously evaluate the performance of entity resolution methods in applications. Second, we discuss potential directions regarding scaling Bayesian entity resolution methods. Finally, we discuss privacy issues surrounding the use of entity resolution.

\paragraph{Evaluating Entity Resolution Performance.}
Despite methodological advances, evaluating the performance of entity resolution remains a challenge for a number of reasons. \cite{murray2020discussion} expressed concerns regarding over-reliance on simple (toy) datasets that may not be representative of real applications, as this could potentially lead methodological research astray.
As a starting point, we review in Section A.2 of the supplementary materials some public datasets that can be used for comparisons/evaluations.  However, given the wide range of fields of application of entity resolution, these datasets are comparatively few in number. We stress that when using ``benchmark data sets," it is crucial that researchers note the number of records under consideration, the level of noise in the data, its overall quality, and the reliability of the unique identifiers used for performance evaluation. In addition, one should not solely rely on toy data sets, but one should perform carefully thought out simulation studies in order to {understand} robustness {to} model misspecification to provide practitioners with a guide for using their method. In addition, extreme care should be taken regarding sensitivity of tuning parameters in proposed methods and sensitivity of the evaluated performance on choices of such parameters. Finally, case studies should be considered if possible as this gives one an idea of how proposed methods work for ``data in the wild." 

In addition, many researchers advocate the use of expert-labeled data to help train entity resolution model and to evaluate their performance in applications. However, care should be taken as labeling errors and sampling procedures may introduce bias into estimates. Effectively eliciting expert-labeled data while accounting for such sources of bias is an active area of research, and one that we consider to be its own field given the complexities involved.

\paragraph{Scaling Entity Resolution.}

Bayesian entity resolution algorithms have been successful in scaling to large datasets, as illustrated by \cite{mcveigh2019scaling} and \cite{marchant_distributed_2019}. \cite{mcveigh2019scaling} has scaled a Bayesian Fellegi-Sunter approach to roughly 57 millions of records using so-called post-hoc blocks. The approach of \cite{marchant_distributed_2019} is quite different as blocking and entity resolution are jointly modeled in a Bayesian framework, allowing for uncertainty quantification about both parts of the pipeline. The authors scaled to roughly one million records using distributed computing. Further research is needed in this area to scale to larger datasets, such as census-size data or industrial sized data sets, while accounting for uncertainties encountered at \emph{all stages} of the entity resolution pipeline.


\paragraph{Privacy Issues.}

Entity resolution is fundamentally antithetic to data privacy. Instead, it is about gaining information about social entities through the integration of diverse databases. This raises ethical and legal questions for users of entity resolution as well as important privacy considerations \citep{lane2014privacy}. In particular, as more data is being collected, stored, analyzed and shared across multiple domains, disclosure risks associated with (even anonymized) data releases become serious. For example, \cite{Narayanan2008} showed how a simple record linkage algorithm could be used to de-anonymize Netflix movie rankings data through the use of public IMDb profiles. \cite{Sweeney2002} used public voter registration data to de-anonymize a health insurance dataset, in order to showcase the need for stronger privacy measures. These are examples of \textit{linkage attacks}, where an adversary uses background knowledge (such as voter registration files) to de-anonymize data or to gain information about individuals.

Data releases should therefore be managed through statistical disclosure control (SDC) systems, which aim to balance the utility of released data with privacy protections. To address these competing goals, many SDC techniques have been proposed and implemented such as top-coding, data swapping, data perturbation, and synthetic data generation, each potentially having its own measures of utility and risk properties; more details are those methods can be found in \citep{fienberg10, hundepool2012sdc}. Furthermore, differential privacy \citep{DMNS06} has emerged as a key rigorous definition of privacy. It provides a framework that can inform the design of privacy mechanisms with specified disclosure risks, in the presence of arbitrary external information.

As we have discussed, analyses often require or can benefit from the linkage of multiple databases. However, when databases are held by different organizations and contain private information that cannot be shared across them, record linkage should be done as to ensure that: (1) private information such as quasi-identifiers (name, date of birth, etc) are not disclosed across organizations during the linkage process, and (2) only relevant summaries of the resulting linkage (usually a set of pre-determined attributes of the linked records) are reported as to manage disclosure risks. The achievement of these two goals is the subject of privacy-preserving record linkage (PPRL) \citep{Hall2010, Vatsalan2013, Vatsalan2017}. This is closely related to the problem of {private multi-party data publishing} under a {vertical partitioning} scheme \citep{Jiang2006, Mohammed2011, Mohammed2014, Cheng2020}. While progress has been made on point (1), the privacy implications of post-linkage data releases are difficult to analyze even under mild adversary models \citep{Mohammed2011, Vatsalan2013}. For instance, any organization involved in PPRL could use its own records to attempt de-anonymizing the released linked data. In other scenarios, an adversarial organization could also use PPRL to gain information about particular individuals. Great care should be taken when using PPRL as to ensure that all disclosure risks are properly assessed.

\section*{Supplementary Materials}

Please refer to \textit{Supplementary Materials for ``(Almost) All of Entity Resolution''} to access the following contents.

\begin{description}
    \item[Open Source Software:] Open source entity resolution software is reviewed in Section A.1 of the supplementary materials.
    \item[Data Sets:] Entity resolution data sets are reviewed in Section A.2 of the supplementary materials.
\end{description}


\title{\bf Supplementary Materials for\\``(Almost) All of Entity Resolution''}
\maketitle

\appendix
\section{Open Source Software and Data Sets}
In this supplement, we review open source entity resolution software (Appendix \ref{app:open-source} and entity resolution data sets (Appendix \ref{app:datasets}).
\label{sec:appendix}

\subsection{Open Source Software}
\label{app:open-source}

This section reviews open source entity resolution software. We focus on libraries available in {R} or {Python} software packages, however, we cover a few recent 
packages that are available in Julia and Apache Spark. Other software is reviewed in \cite{Kopcke2010, christen_data_2012, Christophides2019}.

\paragraph{Available in Python.}

The python library \texttt{dedup} \citep{Gregg2015}, available on PyPI and on GitHub, implements the Fellegi-Sunter framework together with active learning to select threshold weights. Based on this probabilistic record linkage step, it allows clustering records in coreferent groups using hierarchical agglomerative clustering with a centroid linkage. It is widely used, well documented and well maintained. The library \texttt{recordlinkage} \citep{deBruin2016}, available on PyPI and on GitHub, implements the Fellegi-Sunter framework, $k$-means clustering and a number of fully supervised classifiers (logistic regression, support vector machines, etc). The Freely Extensible Biomedical Record Linkage \texttt{FEBRL} library \citep{Christen08febrl}, available on SourceForge, provides a graphical user interface and implements the Fellegi-Sunter framework as well as supervised classifiers and clustering algorithms. Also, the library \texttt{py-entitymatching} \citep{Yash2019}, available on PyPI and GitHub, provides tools to facilitate the development of entity resolution models. It implements rule-based systems as well as a number of supervised machine learning classifiers. Finally,
the package \texttt{fasthash} available on Github implements the work of \cite{chen2018unique}.

\paragraph{Available in {R}.}

The \texttt{RecordLinkage} package on {CRAN} \citep{sariyar2010recordlinkage} implements the Fellegi-Sunter framework and a number of supervised algorithms (logistic regression, support vector machines, random forests, and others). It also contains the two data sets \texttt{RLdata500} and \texttt{RLdata10000} which have been widely used in the literature as benchmark data sets. \cite{enamorado_using_2019} extended the work of \cite{fellegi_theory_1969} and provided efficient open source software on {CRAN} and GitHub known as \texttt{fastLink}. This package supports record linkage, but not de-duplication at this time. The \texttt{blink} package on {CRAN} and GitHub implements the work of \cite{steorts_entity_2015}. The \texttt{representr} package on GitHub implements the work of \cite{kaplan} for canonicalization. 

\paragraph{Available in Julia.}
\cite{mcveigh2019scaling} provide a Julia package to perform blocking and  Bayesian Fellegi-Sunter  called \texttt{BayesianRecordLinkage.jl} on GitHub.

\paragraph{Available in Apache Spark.}
\cite{marchant_distributed_2019} provide a joint blocking and entity resolution package on GitHub, which is provided in Apache Spark with a Java and Scala back-end.


\clearpage
\newpage

\subsection{Entity Resolution Data Sets}
\label{app:datasets}
In this section, we review entity resolution data sets that are publicly available. 

\subsubsection{Synthetic data sets}
First we review synthetic data sets that are publicly available. For all of these data sets, a unique identifier is available to evaluate entity resolution performance.

\paragraph{RLdata.} This contains the \texttt{RLdata500} and \texttt{RLdata10000} synthetic data sets from the \texttt{RecordLinkage} package in \texttt{R} with a total of 500 and 10,000 total records and 10 percent duplication. 
Feature information available is first and last name and full date of birth. 
Both data sets contain a very small amount of distortion in the features. 


\paragraph{GeCo Tool.} One is able to create a synthetic data set using the \texttt{GeCo Tool} \citep{Tran2013}, where features can consist of first name, last name, and birth date. Distortions can be included as to emulate the effect of optical character recognition, keyboard errors, phonetic errors, and common misspellings.

\paragraph{FEBRL.}
The \texttt{FEBRL} data sets \citep{Christen08febrl} consist of comparison patterns from an epidemiological cancer study in Germany (\url{https://recordlinkage.readthedocs.io/en/latest/ref-datasets.html}). The FEBRL2 data set contains 5000 records (4000 originals and 1000 duplicates), with a Poisson distribution of duplicate records truncated at 5. The FEBRL3 data set contains 5000 records (2000 originals and 3000 duplicates), with a maximum of 5 duplicates based on one original record (and a Zipf distribution of duplicate records). The FEBRL1 data set is mostly likely not recommended for record linkage publications or as a benchmark data set given that it is small and the maximum cluster size is 1, making it quite unrealistic.

\paragraph{ABSEmployee.}
The \texttt{ABSEmployee} synthetic data set was constructed to mimic real data from the Australian Bureau of Statistics (ABS), which cannot be released due to privacy reasons. \cite{marchant_distributed_2019} simulated three data sources from the ABS that results in 666,000 total records, with 400,000 unique entities. The three data sources are a supplementary survey of permanent employees (source A), a supplementary survey of all employees (source B), and a census of all employees (source C). The size of source A is 120,000; the size of source B is 180,000; the size of source C is  360,000. Duplication occurs across and within the three data sources.

Feature information available is statistical area level of the employee, mesh block, birth day, birth year, gender (binary), industry, whether employment is on a casual basis (binary), whether employment is full-time, hours worked per week, payrate, average weekly earnings. In all sources, there are missing variables, which are explained further at \url{https://github.com/cleanzr/dblink-experiments/tree/master/data}. 

\subsubsection{Real Data Sets (Publicly Available)}


In this section, we review data sets from the literature which arise from real applications and which are publicly available. For all of these data sets, except for the 1901 and 1911 Irish Census, unique identifiers are available to evaluate entity resolution performance. However, the reliability of these unique identifiers vary. In some cases, these unique identifiers were obtained as the result of extensive record linkage efforts involving expert clerical review of the data. In other cases, the unique identifiers were obtained using external information which is not provided in these data sets.

\paragraph{Cora.}
The \texttt{cora} data set consists of citations and is hosted on the \texttt{RIDDLE} repository \citep{bilenko2006riddle}. Features include title, author, and year of publication. This data set needs some pre-processing steps before a record linkage method can be applied, such as removing punctuation. 

\paragraph{SHIW.}
The Italian Survey on Household and Wealth (FWIW) is a sample survey conducted by the Bank of Italy every two years. The 2010 survey covers 7,951 households composed of 19,836 individuals. The 2008 survey covers 19,907 individuals and 13,266 individuals. The entire survey covers all twenty regions of Italy. Features available are categorical due to privacy reasons, and are the following: year of birth, working status, employment status, branch of activity, town size, geographical area of birth, sex, whether or not Italian national, and highest educational level obtained.
A script that downloads the data set can be found here: \url{https://github.com/ngmarchant/shiw}.

\paragraph{NLTCS.}
The National Long Term Care Survey (NLTCS), a longitudinal study of the health of elderly (65+) individuals (\url{http://www.nltcs.aas.duke.edu/}). The NLTCS was conducted approximately every six years, with each wave containing roughly 20,000 individuals. Unfortunately, only three waves are appropriate for record linkage due to issues with the survey design. Thus, only a subset can be utilized, which are waves 1982, 1989 and 1994. The features available for linking are all categorical and are: gender (SEX), full date of birth (DOB), location of the patient (STATE) and office location of the physician (REGOFF). The provided unique identifier is based upon the social security number.

\paragraph{CD.}
The \texttt{CD} data set includes information about 9,763 CDs randomly extracted from freeDB.\footnote{This data set can be found at \url{https://hpi.de/naumann/projects/repeatability/datasets/cd-datasets.html}.} There are a total of 299 duplicate records. Attribute information consists of 106 total features such as artist name, title, genre, among others. 

\paragraph{Restaurant.}

The \texttt{Restaurant} data set contains 864 restaurant records collected from Fodor's and Zagat's restaurant guides.\footnote{This data set was originally provided by Sheila Tejada, and was downloaded from \url{http://www.cs.utexas.edu/users/ml/riddle/data.html}.}  There are a total of 112 duplicate records. Attribute information contains name, address, city, and cuisine. 

\paragraph{NCSBE.}
The North Carolina State Board of Elections (NCSBE) releases an online publication of North Carolina voter registration snapshot data. Records are updated temporally, resulting in voters being duplicated within this data set. While the NCSBE provides each voter with an identifier in each of the snapshots, they do not provide any public information regarding how duplicate records are removed. In addition, the reliability of the NCSBE ``unique'' voter identifiers has been recently been questioned \citep{wortman2019record}. Feature information consists of first and last name, age, gender, race, place of birth, age, political affiliation, telephone number, and full address. 




\paragraph{USPTO.} In 2015, PatentsView\footnote{See \url{https://www.patentsview.org/}.} organized a competition aiming to develop an inventor disambiguation algorithm for the USPTO patents records. Five data sets of inventor-disambiguated patent records were provided as training data to help develop proposed algorithms and can be downloaded from \url{https://community.patentsview.org/workshop-2015}. A research data set of all patents which PatentsView has now disambiguated using the winning algorithm of the 2015 competition can be downloaded from \url{https://www.patentsview.org/download}.




\paragraph{SDS.}

The \texttt{Social Diagnosis Survey (SDS)} is a project that supports diagnosis work derived from institutional indicators of quality of life in households in Poland (anyone older than sixteen years of age). The first sample was taken in the year 2000 and the same households were revisited roughly every two years afterwards as a followup survey, up to 2015. The complete data can be found at \url{http://www.diagnoza.com/index-en.html}. The entire SDS consists of 
41,227 unique records of individual members of households that participated in the survey in at least one of the years 2011, 2013, and 2015. Individuals can be duplicated longitudinally across these three waves, however, duplication cannot occur within a wave. Attribute information consists of the following categorical information: sex, full date of birth, province of residence, and education level. 

\paragraph{SIPP.}
The \texttt{Survey of Income and Program Participation (SIPP)} is a longitudinal survey that collects information about the income and participation in federal, state, and local programs of individuals and households in the United States \citep{SIPP}. The SIPP is administered every few years in panels, where sampled individuals within each panel are divided into four rotation groups (subsamples of roughly equal size). One rotation group is interviewed each month such that a wave of the survey consists of a four-month cycle of interviews. The data is publicly available from the Census Bureau website at \url{https://www.census.gov/programs-surveys/sipp/data/datasets.html}. Feature information available is sex, year and month of birth, race, and state of residence. 


\paragraph{1901 and 1911 Irish Census.}
Each household in Ireland was expected to complete what is referred to as Form A, which records the names of all the individuals in the home on the night of 31st March 1901.\footnote{Enumerators, typically local police constables, went from house to house collecting the census forms from literate families, and filled in the required information for those heads of household who could not read and write. These census forms were signed by both the enumerator and the head of household, no matter who had filled in the required information.}
Form A contains all members that resided in a household and any visitors (that stayed the night). Attribute information recorded is full name, age, gender, relationship to head of household, religion, occupation, marital status, county of birth (unless born abroad, in which case only the country was recorded), ability to read or write,
ability to speak Irish, English, both, or none. There were other forms that were required for businesses, ships, and other types of reasons, which can be found at \url{https://www.irish-genealogy-toolkit.com/census-forms.html}. 

The census of 1911 for Form A was slightly different than that of 1901. The household completed and signed this form. One additional attribute was disability status. Three additional questions were asked of married women: the number of years with their current husband, the number of children (born alive), and the number of children still living. Similar to the census of 1901, there were other forms required for businesses, ships, etc. For those interested in using the data, all thirty-two Irish counties for 1901 and 1911 can be found at\\ \url{http://www.census.nationalarchives.ie/}.

\subsubsection{Real Sata Sets (Private)}
In this section, we review data sets that are not available in the public domain, but 
have an important place in the literature.

\paragraph{El Salvador -- ER-TL and CDHES.}

As already mentioned, El Salvador experienced a civil was from 1980 --- 1991. We have discussed in Section~1.1 the United Nations Truth Commission (UNTC) data set. Additionally, there were two other sources that collected information --  El Rescate - Tutela Legal (ER-TL) and the Salvadoran Human Rights Commission (CDHES). 

El Rescate, based out of Los Angelos, CA, developed a database on human rights abuses, where the information was digitized from reports that were published by \emph{Tutela Legal} of the Archdiocese of San Salvador. The information published came from individuals that came to \emph{Tutela Legal}'s office in San Salvador in order to make denunciations. According to \cite{howland2008rescate}, \emph{Tutela Legal} performed interviews, checked credibility of  denunciations and testimonials, and compared any denunciations with existing records to avoid duplications in their databases. Unfortunately due to the nature of data collection, it's not possible to publish all denunciations as it is possible that some were not reported due to restrictions of access to other areas or occurrences of under-reporting. The size of this database is 4,420 records. 

CDHES collected data between 1979 and 1991 according to \cite{ball2000salvadoran}, and collected more than 9,000 testimonials that were recorded in writing. In 1992, these recordings were digitized and a database was created to summarize a summary of all reported violations. The attributes present in both databases are: given and family names of the victim, full date of death, and region of death. The size of this database is 1,320 records. 

Neither of these data sets is publicly available to our knowledge. Further information regarding these data sets is summarized in \cite{sadinle_bayesian_2017}.

\paragraph{Syria.}

The \texttt{Syria} data set comprises data from the Syrian conflict, which covers the same time period, namely, March 2011 -- April 2014. This data set is not entirely publicly available and was provided by the Human Rights Data Analysis Group (HRDAG). The respective data sets come from the Violation Documentation Centre (VDC), Syrian Center for Statistics and Research (CSR-SY), Syrian Network for Human Rights (SNHR), and Syria Shuhada website (SS). Each database lists a different number of recorded victims killed in the Syrian conflict, along with available identifying information including full Arabic name, date of death, death location, and gender.\footnote{These databases include documented identifiable victims and not those who are missing in the conflict. Hence, any estimate reported is only a lower bound on the true number of victims.} For more information on this data set, we refer to \cite{Price2014, Sadosky2015, chen2018unique, tancredi2018unified}.

%
%
%
%

\paragraph{Decennial Census and Administrative Records.}

We have introduced the decennial census data problem in Section~1.1: every ten years, when the U.S.\ Census Bureau 
must enumerate the population in each state as mandated under the U.S.\ Constitution, Article~I,~Section 2. Many individuals elect not to fill out census forms, 
which results in them not being counted in the enumeration. 
Other individuals may be counted multiple times due to duplicate responses. 
For example, students attending universities or private schools (living in group quarters) are often double counted as they are legally required to be counted by their university\slash school, while also being counted by their parents\slash guardians as part of a household. 
%
One way to improve coverage is to leverage administrative data, such as the 
Social Security Administration's Numerical Identification System 
(Numident). The Numident is the Social Security Administration's 
computer database file of an abstract of the information contained in an 
application for a U.S.\ Social Security number.  The following attributes are available for each data set: last name, date of birth, gender, and zip code, which are protected under Title 13, and thus, this data cannot be shared. \cite{marchant_distributed_2019} recently performed a case study on a subset of this data for the state of Wyoming, where the results are described in their paper.

\paragraph{ANES and PVF.}

The American National Election Studies (ANES) is a national survey of voters in the United States (U.S.) that is conducted before and after every presidential election.
The target population of the ANES
are U.S. citizens eligible to vote.
Since 2008, the ANES is 
carried out via two interview modes --- 
traditional face-to-face interviews and
more common internet interviews. 
The face-to-face component of
the ANES is a multistage stratified cluster sample 
of residential addresses, which due to financial constraints 
does not include Alaska and Hawaii. 
The internet component of the ANES is a random sample of residential
addresses in the U.S. states and the
District of Columbia. In 2016, out of 4,271 ANES respondents, 
1,181 were face-to-face and 3,090 were internet respondents. 
The corresponding attrition rate from the pre-election to post-election surveys was
10\% for the face-to-face and 16\% for the internet respondents. For details
on the sampling design see \citet{anes16:meth}.

L2, a leading national nonpartisan firm, will provide to some researchers a nationwide voter file from the 2016 presidential election of over 180 million records. In this file, all states have updated their voter files by including information about the 2016 election. It is possible that through routine data cleaning by states and/or L2 that some individuals who voted in the election may have been removed because they either have deceased or
moved. As a result, the L2 voter file has a total of 131 million voters who cast their ballots, whereas, according to the United States Election Projection, approximately 136.7 million individuals voted in the election. In addition, the L2 voter file does not contain overseas voters, which reduces this data set by roughly 5 million voters, and the turn out rate by slightly more than one percentage point. This data set is referred to as the Presidential Voter File (PVF).

In \cite{enamorado_using_2019}, the authors were motivated to understand what drives the large gap between self-reported turnout in the 2016 ANES data set and the actual turnout among the voting-eligible population in 2016. To understand this further, one must compare the self-reported behavior (2016 ANES) with actual behavior (the 2016 PVF). Thus, this requires record linkage of these two databases. Unfortunately, neither are publicly available.\footnote{There cannot be duplications within the ANES and PVF files, as it is reasonable to assume that an individual responds or votes only one time.}

\paragraph{California Great Registers.}
Starting in 1900, each country in California (CA) printed and bound voter lists in each election year, which contained the following feature information of each voter: name, address, party registration, and occupation \citep{spahn2017before, mcveigh2019scaling}. This became known as the \texttt{California Great Registers} data set, and was used as the county's form of book keeping on election day. These original voter lists have now been digitized using ancestry.com and optical character recognition, however, this can cause errors in the data. The entire data set can be viewed as a panel data set, where it may be possible to track partisan change during certain time periods. This data set spans 1908 --- 1968. It is possible to potentially match voters from this time period with individuals from three decennial censuses from 1920, 1930, and 1940, which are publicly available. To our knowledge, the California registers database is not publicly available. Together, the three decennial censuses and the \texttt{California Great Registers} data set combine to form a data set of 57 million records of Californians. 


%


\bibliographystyle{jasa}
\bibliography{er-review}

\end{document}